\documentclass[epj]{svjour}
\usepackage{graphics}
\usepackage[utf8]{inputenc}
\begin{document}
\title{Information entropy of classical versus explosive percolation}
\author{T. M. Vieira\inst{1} \and G. M. Viswanathan\inst{1,2} \and L. R. da Silva\inst{1,2}
}                     
%
%
\institute{Departamento de Física, Universidade Federal do Rio Grande do Norte, 59078-970 Natal, Rio Grande do Norte, Brazil \and National Institute of Science and Technology of Complex Systems, 
Universidade Federal do Rio Grande do Norte, 59078-970 Natal, Rio Grande do Norte, Brazil}
\date{Received: date / Revised version: date}
%
\abstract{
We study the Shannon entropy of the cluster size distribution in classical as well as explosive percolation, in order to estimate the uncertainty in the sizes of randomly chosen clusters. At the critical point the cluster size distribution is a power-law, i.e. there are clusters of all sizes, so one expects the information entropy to attain a maximum. As expected, our results show that the entropy attains a maximum at this point for classical percolation. Surprisingly, for explosive percolation the maximum entropy does not match the critical point. Moreover, we show that it is possible determine the critical point without using the conventional order parameter, just analysing the entropy's derivatives.
%
} 
\maketitle
\section{Introduction}

\begin{figure}
\resizebox{0.5\textwidth}{!}{ \includegraphics{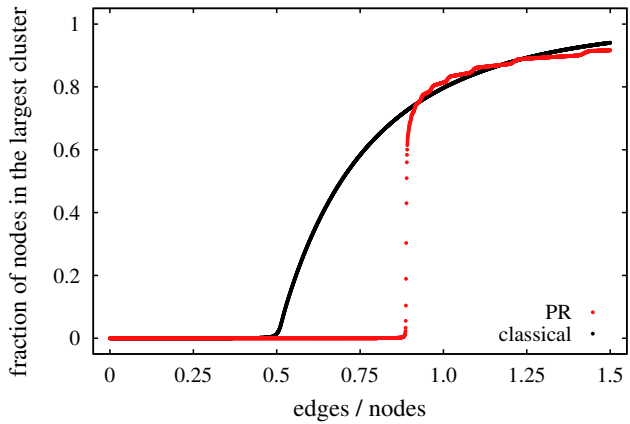} }
\caption{\label{plot-01}(Color online) Illustration of percolation phase transitions in random networks (non-lattice). Note the smooth growing curve for the classical percolation and the step-shaped curve for the explosive percolation (product rule --- PR). The plot shows the fraction of nodes in the networks' largest cluster vs. the density of edges, i.e. the ratio between the number of edges and the number of nodes.  These results, and all others hereafter, except when it's explicitly mentioned, are based on simulation of 500 percolation transitions of each type over networks with $3 \times 10^5$ nodes.}
\end{figure}

Percolation~\cite{stauffer-1994,saberi-2015} is used to model diverse phenomena, ranging from porous media to social interactions. There is a well-known phase transition associated with percolation. At the critical point, the giant cluster emerges, comparable in size to the entire network. The classical network model developed by Erdös and Rényi~\cite{erdos-1960} (random network) has a smooth continuous phase transition, as shown in Fig.~\ref{plot-01}. In this model edges are randomly arranged between the nodes in the network. It was believed until recently that all percolation transitions were continuous. However, the discovery of explosive percolation~\cite{achlioptas-2009} led to questions regarding the continuous nature of the percolation phase  transition~\cite{costa-2010,grassberger-2011,riordan-2011,cho-2011,%
lee-2011,tian-2012,costa-2014,costa-2014-2}. 

The key idea behind explosive percolation is to add edges in a manner to delay the onset of the percolation transition. A specific choice is made before the addition of each new edge. This choice process has the aim of delaying the formation of the giant cluster. As a consequence, it grows very suddenly at birth, leading to an abrupt phase transition. The first proposed mechanism was called the product rule (PR)~\cite{achlioptas-2009}, which works as follows. Two node pairs are selected and an edge is placed between the pair whose product of the number of nodes in the connecting clusters is the smallest. The explosive percolation caused by the PR is shown in Fig.~\ref{plot-01}. 
Actually researches agree that the explosive percolation transition is continuous, but with unusual characteristics~\cite{costa-2014-2}. In fact, explosive percolation is not yet completely understood and continues to be an area of intense research~\cite{araujo-2014}. 

Looking for a method of studying phase transitions of percolating systems without the explicit use of order parameter, we analyze the Shannon entropy associated with the cluster size probability distribution. 
By doing so, we note that it's possible to determine the critical point through the entropy's derivatives. In addition, we find a previously unknown (and unexpected) property of explosive percolation: in contrast with classical percolation, its critical point does not correspond to a maximum of the Shannon entropy of the cluster size distribution.

\section{Information entropy of the cluster size probability distribution}

We propose a new way of observing and studying percolation phase transitions. Our method is capable of determining the transition point even without the specific study of the order parameter (relative size of the network's largest cluster for classical percolation on random networks~\cite{stauffer-1994,costa-2014}). Although others mechanisms towards abrupt percolation transition have been proposed~\cite{dsouza-2010,araujo-2010,manna-2011}, in this article we focus only on: (i) the product rule (PR) and (ii) the rule developed by da Costa~\textit{et al.}~\cite{costa-2010,costa-2014} (dCR). The explosive percolations resulting from these two mechanisms and the classical percolation were simulated and we analyzed them through the alternative approach described below.

Our new method for observe percolation phase transitions is based on information theory. We use the Shannon entropy~\cite{shannon-1948}, that is defined as
\begin{equation} 
H = -K \sum _i \left( p_i \log_2 p_i \right) , 
\label{eq-shannon-entropy}
\end{equation} 
in which the set $\{p_i\}$ is a probability distribution and $K$ is a constant.  It is a function whose output is a type of measure of information contained in the probability distribution~$\{p_i\}$. While we arrange edges in the network, clusters are formed so that the cluster size probability distribution changes during this process. The cluster size distribution has the following meaning: at a given moment $t$ the network has $n_t(1)$ clusters with size 1 (single nodes), $n_t(2)$ clusters with size 2, and so on. After normalization, we have the probability distribution. Thus, at each moment, this distribution can be used to characterize the current state of the network. We employ the Shannon entropy to analyze the information content of this distribution.

The Shannon entropy is such that the more flat or uniform the probability distribution (that is, tending to the equiprobability), the higher the entropy $H$. Thus when there are only single-node clusters, we obtain $H=0$. There is no uncertainty about the size of the clusters (i.e. they all have size 1). As edges are added to the network, the entropy starts to increase because of the emergence of new clusters with different sizes.  However, there is a moment when the network is populated by several clusters of various sizes and the cluster size distribution attains its widest form, so $H$ reaches a maximum. At the critical point the system becomes scale-free and a cluster with size comparable to the size of the network emerges. This moment announces the giant cluster onset and the percolation transition. Thereafter, the giant cluster starts to absorb other clusters, thereby decreasing the amount of clusters and also the variety of different sizes. Therefore, one expects that $H$ decreases after the critical point. This is basically the dynamics observed for the classical percolation, while that resulting from the explosive rules has more ingredients.

\begin{figure}[t]
\resizebox{0.5\textwidth}{!}{ \includegraphics{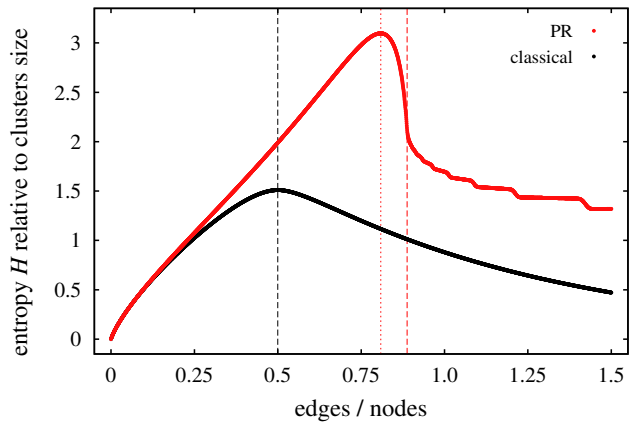} }
\caption{\label{plot-02}(Color online) Shannon entropy (Eq.~\ref{eq-shannon-entropy} with $K=1$) of the cluster size probability distribution vs. density of edges. The red data are the result for networks whose edges are added through PR, while the black data are the result for networks undergoing the classical percolation. The dashed lines mark in which value for the density of edges the percolation transition occurs: accurately at 0.5 for classical percolation (black) and around 0.8885 for explosive percolation (red). We observe that the maximum value for the entropy matches the classical percolation's critical point. Looking at explosive percolation data, the maximum entropy (red dotted line) occurs after that for classical percolation and, more important, the transition doesn't happen at the maximum entropy but afterwards.  The value for the density of edges at the maximum entropy for the explosive case is $0.8085(5)$. The step-shaped pattern observed after the transition point is a manifestation of the finite size of the system and is a characteristic of the PR mechanism.}
\end{figure}

Figure~\ref{plot-02} shows graphically what we have discussed above. 
We have used Eq.~\ref{eq-shannon-entropy} with $K=1$ and repeatedly calculated the entropy after a fixed number of edges has been added to the networks. The result for classical percolation follows the expected behavior and the maximum entropy matches the critical point. However, the result for explosive percolation doesn't. The explosive transition happens only after the addition of yet more edges. Figure~\ref{plot-02} shows that the PR leads the networks to a configuration of maximum entropy, although the giant cluster still has not been born. Between the point of maximum entropy and the critical point the entropy declines rapidly. 

We emphasize that the critical point is marked by a power-law cluster distribution (not shown), which means that there are clusters of all sizes, with scale-free behavior. Nevertheless our simulations show clearly that the critical point does not maximize the entropy. How can this be? We note that the size of the clusters at the maximum entropy configuration is limited, tending to be negligible for networks composed by a very large amount of nodes. Moreover, because the information entropy of cluster size distribution is related with the uncertainty to guess the size of a randomly chosen cluster, we conclude that it is harder to guess the size of a cluster before the transition than at the transition.

\begin{figure}[t]
\resizebox{0.5\textwidth}{!}{ \includegraphics{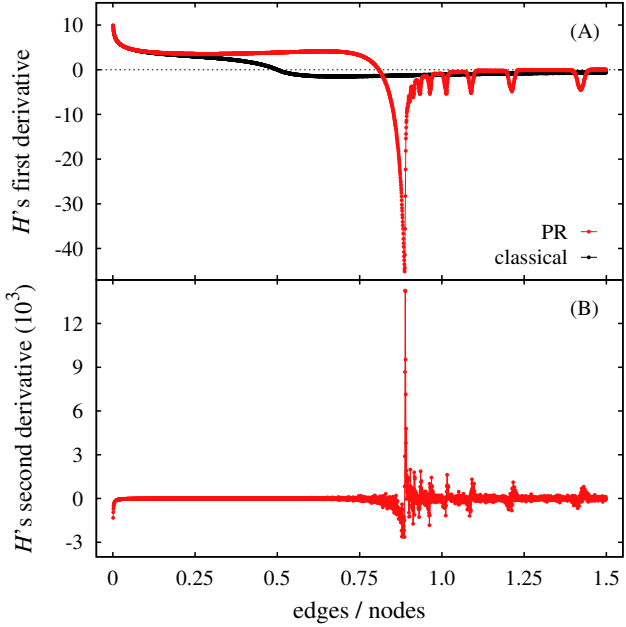} }
\caption{\label{plot-03-ab}(Color online) 
In~(A) the entropy $H$'s first derivative in which there are data for the PR (red) and for classical percolation (black). While the classical percolation data form a smooth curve, the PR data has a striking downward peak just before the critical point. In~(B) the $H$'s second derivative. Because the second derivative doesn't contribute to the analysis of the classical percolation data, we show only this derivative for the PR case. Considering the numerical accuracy of our data, we observe that the occurrence of the maximum value for the second derivative matches the percolation transition~\cite{grassberger-2011} within the error interval. } 
\end{figure}

The data exposed on Fig.~\ref{plot-02} alone is not enough to explain the positioning of the explosive percolation's critical point based on the Shannon entropy analysis. Nevertheless, we note that the uncertainty falls quickly before the transition. So, we calculate the entropy's first and second derivatives to see how they behave in the vicinity of the transition. The numerical derivatives are shown in Fig.~\ref{plot-03-ab} and were obtained as follows. Because in our simulations the density of edges varies uniformly, we represent the step between each subsequent addition of edges by $\Delta$ (constant). With $H_i$ representing the entropy after the \mbox{$i$-th} addition of edges (i.e. at density $i \times \Delta$), we calculate the derivatives using \cite{fornberg-1988} 
\begin{equation}
H'_i = \frac{1}{\Delta}
\left[ {\frac{1}{12} H_{i-2} - \frac{2}{3} H_{i-1} + \frac{2}{3} H_{i+1} - \frac{1}{12} H_{i+2}} \right]
\label{eq-first-derivative}
\end{equation}
for the first derivative and 
\begin{equation}
H''_i = \frac{1}{\Delta ^2}  
\left[{-\frac{1}{12} H_{i-2} + \frac{4}{3} H_{i-1} - \frac{5}{2} H_i + \frac{4}{3} H_{i+1} - \frac{1}{12} H_{i+2}} \right]
\label{eq-second-derivative}
\end{equation}
for the second derivative. 
Such as expected, the curve relative to the classical percolation is smooth (hence the second derivative was suppressed). However, the striking feature is the sharpened curves for the explosive case. It's possible to see the explosive behavior being reproduced. 
In~(A) we see a sharp minimum just before the transition. The fast decreasing in the entropy is caused by the agglutination of smaller clusters. This results in clusters somewhat larger, what increases the amount of clusters of different sizes and decreases the probability associated to the majority of them. This dynamics reduces the entropy relative to the cluster size distribution. 
At this stage there are favorable conditions for the emergence of the largest cluster and the PR prevents it by favoring that smaller clusters would coalesce. 
However, there is a moment when arrange edges between different clusters becomes harder and more edges are arranged internally to a same cluster. So, because the cluster size distribution undergoes less changes, the entropy starts to decrease more slowly and the first derivative exhibits a strongly vertical slope. In that moment, it becomes impossible to prevent the largest cluster growth and it initiates to swallow the others. 
In~(B) we see the entropy's second derivative. The occurrence of its maximum value coincides with the percolation transition. Therefore, we conclude that the transition occurs just after the entropy ceases to fall quickly.

\begin{figure}[t]
\resizebox{0.5\textwidth}{!}{ \includegraphics{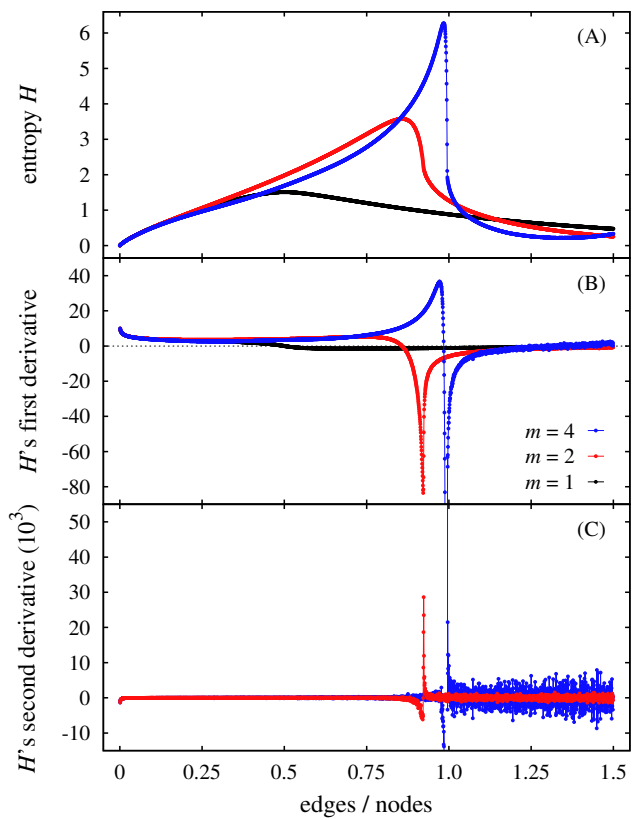} }
\caption{\label{plot-04-abc}(Color online) An alternative way to observe percolation transitions generated via the dCR. These curves show the behavior of the entropy $H$ (A) and its derivatives (B and C) when we increase the density of edges. Ordering from the curve whose maximum is more to the left in~(A): $m=1$ (black), $m=2$ (red) and $m=4$ (blue). The higher is the value of $m$, the more restrictive is the edges addition. While $m=1$ recovers the classical percolation, $m>1$ leads to explosive ones. The occurrences of the maximum entropy and of the critical point are retarded by higher $m$, so the abrupt character becomes more strong. The separation between these two moments is apparent, however they become closer for higher $m$. While the maximum entropy occurs at $0.8575(5)$ and the percolation transition at $0.9235(5)$ for $m=2$, these same points are $0.9840(5)$ and $0.9950(5)$ for $m=4$, respectively. In~(B) and (C), we restricted the data showed for $m=4$, otherwise the data for $m=2$ would not be visible properly. Specifically in plot~C, we suppress the data for $m=1$, because it would be approximately zero all time. }
\end{figure}

The application of our method to the explosive percolation caused by the dCR is shown in Fig.~\ref{plot-04-abc}. This rule imposes that edges are added as follows: (i)~choose two sets of $m$ nodes; (ii)~for each set, select the node in the smallest cluster; and (iii)~put an edge connecting these two nodes. While the case $m=1$ recovers the classical percolation, if $m>1$ we obtain explosive percolations. Such as the explosive percolation via PR, the point of maximum value for the entropy $H$ occurs before the percolation transition. The critical point is indicated by the maximum value for the entropy's second derivative, as before. We can see that the higher is $m$, the more retarded are these instants (maximum entropy and critical point) in comparison to the classical percolation. Furthermore, these points become closer when $m$ is increased, such that they should coincide when the value of $m$ is too high. However, if $m$ is comparable to the size of the network, the edges will be added deterministically. They will always connect the two smallest clusters in the network at every moment. In this case, the cluster size distribution is fully known and calculate its Shannon entropy will not give us more insights about the network structure. Therefore, our approach only applies when there is an amount of randomness involved.

\section{The information entropy analysis is not sensitive to network size}

\begin{figure}[t]
\resizebox{0.5\textwidth}{!}{ \includegraphics{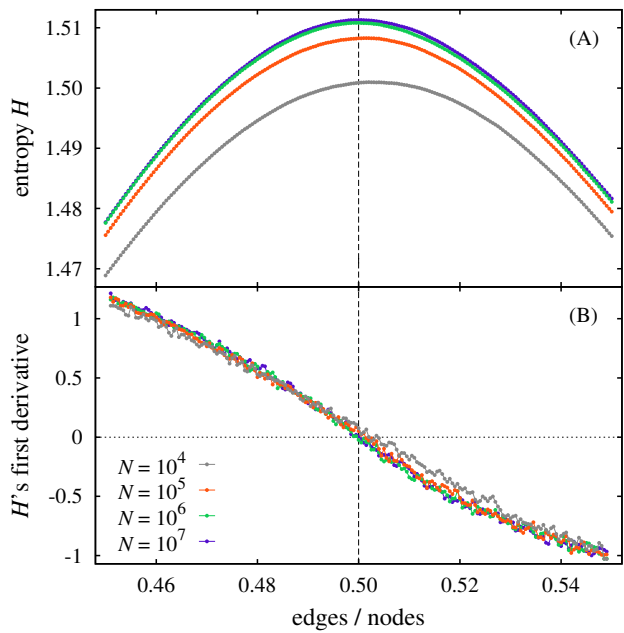} }
\caption{\label{plot-05-ab}(Color online) In~(A) the entropy $H$ for networks with different number of nodes ($N$) undergoing the classical percolation. The plot is restricted to a data interval around the maximum entropy. Ordering from the lowermost curve: $N = 10^4$ (gray), $N = 10^5$ (orange), $N = 10^6$ (green) and $N = 10^7$ (blue). These data show that the entropy of networks with $N>10^6$ collapses over the same limiting curve. However, we note that even networks with $N<10^6$ preserve the same form for the entropy curve, such that the maximum entropy still occurs in the vicinity of critical point. This little gap between the maximum entropy and the critical point for $N < 10^6$ is a visible effect of the finite size of the system. From these data, we infer that the behavior of the entropy in the limit of infinite size does not differ significantly from that for $N = 10^7$. In~(B) the $H$'s first derivative. Despite the observed fluctuations, it's easy to see that the derivative becomes zero at the critical point. The dashed vertical line marks the value of edges density at the maximum entropy and at the critical point at the same time. 
These results are based on simulation of $n$ classical percolation transitions for each $N$ such that $n \times N = 10^8$. }
\end{figure}

\begin{figure}[t]
\resizebox{0.5\textwidth}{!}{ \includegraphics{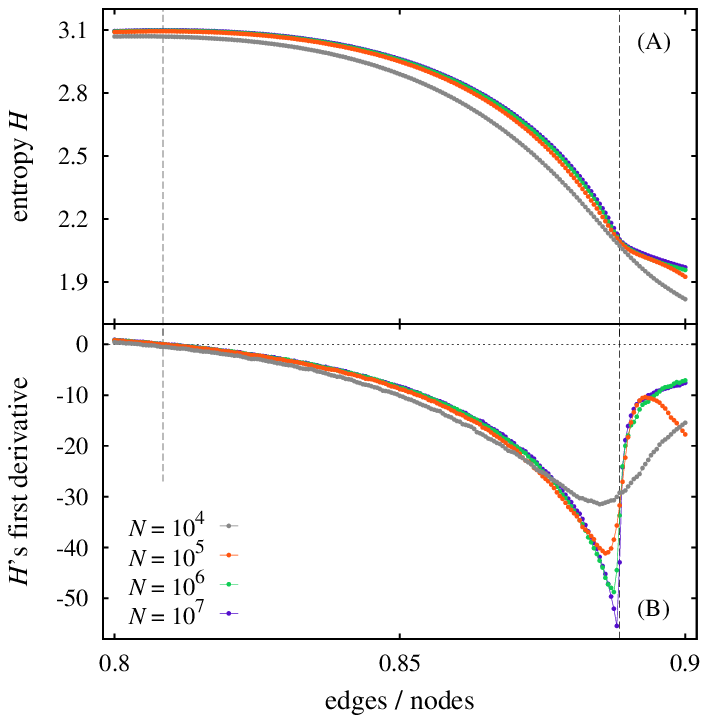} }
\caption{\label{plot-06-ab}(Color online) In~(A) the entropy $H$ for networks with different $N$ and whose edges are added through PR. The plot is restricted to a data interval including the maximum entropy and the critical point. 
The color scheme and the relative position of the curves are analogous to that of Fig.~\ref{plot-05-ab}. 
Only the data for networks with $N=10^4$ do not collapse over the same curve. 
Considering this, we infer that there is a limiting curve corresponding to the behavior of the entropy when the networks size tends to the limit of infinite size and we suppose that this curve does not differ significantly from that for $N = 10^7$. 
In~(B) the $H$'s first derivative. These data reveal that the occurrence of the minimum for the first derivative is shifted towards the critical point while $N$ increases. For $N=10^7$ we find that this minimum for the first derivative and the maximum value for the second derivative (not shown) coincide, and they occur at the percolation transition. The decline of the $H$'s first derivative for $N = 10^5$ after the critical point is a manifestation of the finite size of the system. The left dashed line marks the value of edges density in which $H$ is maximum and the right dashed line marks the critical point. 
These results are based on simulation of $n$ PR percolation transitions for each $N$ such that $n \times N =  10^8$. 
}
\end{figure}

Questions may be raised about what happens with the maximum entropy when the number of nodes in the networks increase. Will the entropy related to classical percolation keeps its smooth appearance when the network size grows? Would it be the mismatch between maximum entropy and critical point only an effect of the finite size of the system? In order to answer such questions, we apply our method to networks composed by different amounts of nodes. 
We restrict this study to classical and PR percolations and the results for them are shown in Figs.~\ref{plot-05-ab} and \ref{plot-06-ab}, respectively. 
In Table~\ref{table-1}, we compile data indicating that the maximum entropy point and the critical point converge to well-determined values when the amount of nodes is increased. In turn, Table~\ref{table-2} contains the relative difference between the value of maximum entropy for growing sizes. It underlies our argument on the convergence of entropy toward a limiting behavior while we grow the size of the network systematically, as described below.

The analysis of the data relative to classical percolation is straightforward, because we need to focus essentially on one point of the graph. 
Figure~\ref{plot-05-ab} shows in (A) the curves getting closer while the size ($N$) of the system is increased, so that the two topmost curves are virtually coincident. 
This pattern, summarized in Table~\ref{table-2}, allows us infer that there is a limiting curve for which all others tend when $N$ increases towards infinite. Moreover, we expect that this limiting curve is located very close to that for $N = 10^7$. In~(B), the $H$'s first derivative data indicate that the curves have the same aspect, so that none of them is sharper around the maximum entropy point than the others. In spite of fluctuations, it's easy to see that the value of the derivative at each point is the same for the studied sizes, being only slightly different to the smaller one ($N = 10^4$). 

In the PR case, the analysis must be more careful, since the maximum entropy and the critical point do not match. In this way, observing Fig.~\ref{plot-06-ab}, we note that the maximum entropy and the critical point remain well-determined graphically when the size of the networks varies. Furthermore, we see that the gap between them does not change while $N$ increases, such as the data in Table~\ref{table-1} demonstrate. 
The curves in~(A) reveal that our method is not sensitive to the network size, since it is greater than a given minimum size, which we estimate to be around $10^5$ nodes. Although effects of finite size of the networks may be observed, especially after the percolation transition, they do not affect the positioning of the maximum entropy point and the determination of the critical point. Indeed, the data for sizes above this minimum size indicate a collapse on the same limiting curve (see Table~\ref{table-2}). Thus, we infer that at the limit in which the system has an infinite size, the shape of the curve for the entropy $H$ will not change from this limiting curve. 
In~(B), we show that while $N$ increases, the first derivative becomes sharper and its minimum tends to coincide with the percolation transition. In fact, we point out that the gap between the minimum of the first derivative and the maximum of the second derivative disappears when the system size becomes too large. 

\begin{table}[t]
\caption{\label{table-1}
Values of edges density corresponding to the maximum entropy (ME) and to the critical point (CP) in networks with different number of nodes ($N$), compiled from the data plotted in Figs.~\ref{plot-05-ab} and \ref{plot-06-ab}. There is a minimum size above of which these values coincide.}
\begin{center}
\begin{tabular}{*4c}
\hline\noalign{\smallskip}
{} & Classical percolation & \multicolumn{2}{c}{Explosive percolation (PR)} \\
$N$     &   ME and CP    &       ME      &       CP      \\
\noalign{\smallskip}\noalign{\smallskip}
$10^4$  &   0.5020(5)    &   0.8045(5)    &    0.8880(5)  \\
$10^5$  &   0.5015(5)    &   0.8085(5)    &    0.8890(5)  \\
$10^6$  &   0.5000(5)    &   0.8085(5)    &    0.8885(5)  \\
$10^7$  &   0.5000(5)    &   0.8085(5)    &    0.8885(5)  \\
\noalign{\smallskip}\hline
\end{tabular}
\end{center}
\end{table}

\begin{table}[t]
\caption{\label{table-2}
Relative difference between the value of maximum entropy for two consecutive curves in Figs.~\ref{plot-05-ab} and \ref{plot-06-ab}. These data show the convergence towards a limiting value. }
\begin{center}
\begin{tabular}{ccc}
\hline\noalign{\smallskip}
$N$ & Classical & Explosive (PR) \\
\noalign{\smallskip}\noalign{\smallskip}
$10^4 \to 10^5$  &  0.49\%  &  0.78\%  \\
$10^5 \to 10^6$  &  0.17\%  &  0.09\%  \\
$10^6 \to 10^7$  &  0.03\%  &  0.07\%  \\
\noalign{\smallskip}\hline
\end{tabular}
\end{center}
\end{table}

\section{Conclusion}

In summary, we have developed a new method for studying percolation phase transitions, which makes use of the Shannon entropy of the cluster size distribution. Although the study of the order parameter is the canonical way to observe phase transitions of percolating systems, some aspects of the network evolution toward the percolation transition are not apparent via this conventional approach. An immediate and unexpected output of our approach is that the explosive percolation transition does not occur at the maximum of entropy. This adds up to other characteristics of explosive percolation contrasting with classical percolation. Although the network is populated by clusters of all sizes at the critical point, there are other configurations with even higher entropy of cluster sizes, if edges are added following explosive mechanisms. We show that it is possible determine the critical point of classical and of explosive percolations on random networks only through the Shannon entropy's derivatives. In addition, our computational analysis demonstrates that this approach has a fast convergence to the expected behavior to a system in the limit of infinite size.  In this context, we hope that our results motivate further studies about percolating systems from non-canonical viewpoints with the aim of increasing our understanding of percolation transitions, both classical as well as explosive.

\vspace{5mm}
\noindent
The authors thank the financial support of CNPq.

\vspace{5mm}
\noindent
All authors contributed equally to the paper.

\bibliography{arxiv-abrupt-v31}

\providecommand{\noopsort}[1]{}\providecommand{\singleletter}[1]{#1}%
\begin{thebibliography}{10}

\bibitem{stauffer-1994}
D.~Stauffer and A.~Aharony, {\em Introduction to Percolation Theory}.
\newblock CRC Press, 1994.

\bibitem{saberi-2015}
A.~A. Saberi {\em Phys. Rep.}, vol.~578, pp.~1--32, 2015.

\bibitem{erdos-1960}
P.~Erdös and A.~Rényi {\em Publ. Math. Inst. Hungar. Acad. Sci.}, vol.~5,
  pp.~17--61, 1960.

\bibitem{achlioptas-2009}
D.~Achlioptas, R.~M. D’Souza, and J.~Spencer {\em Science}, vol.~323,
  p.~1453, 2009.

\bibitem{costa-2010}
R.~A. da~Costa, S.~N. Dorogovtsev, A.~V. Goltsev, and J.~F.~F. Mendes {\em
  Phys. Rev. Lett.}, vol.~105, p.~255701, 2010.

\bibitem{grassberger-2011}
P.~Grassberger, C.~Christensen, G.~Bizhani, S.-W. Son, and M.~Paczuski {\em
  Phys. Rev. Lett.}, vol.~106, p.~225701, 2011.

\bibitem{riordan-2011}
O.~Riordan and L.~Warnke {\em Science}, vol.~333, pp.~322--324, 2011.

\bibitem{cho-2011}
Y.~S. Cho and B.~Kahng {\em Phys. Rev. Lett.}, vol.~107, p.~275703, 2011.

\bibitem{lee-2011}
H.~K. Lee, B.~J. Kim, and H.~Park {\em Phys. Rev. E}, vol.~84, p.~020101, 2011.

\bibitem{tian-2012}
L.~Tian and D.-N. Shi {\em Phys. Lett. A}, vol.~376, pp.~286--289, 2012.

\bibitem{costa-2014}
R.~A. da~Costa, S.~N. Dorogovtsev, A.~V. Goltsev, and J.~F.~F. Mendes {\em
  Phys. Rev. E}, vol.~89, p.~042148, 2014.

\bibitem{costa-2014-2}
R.~A. da~Costa, S.~N. Dorogovtsev, A.~V. Goltsev, and J.~F.~F. Mendes {\em
  Phys. Rev. E}, vol.~90, p.~022145, 2014.

\bibitem{araujo-2014}
N.~Araújo, P.~Grassberger, B.~Kahng, K.~J. Schrenk, and R.~M. Ziff {\em Eur.
  Phys. J. Special Topics}, vol.~223, pp.~2307--2321, 2014.

\bibitem{dsouza-2010}
R.~M. D’Souza and M.~Mitzenmacher {\em Phys. Rev. Lett.}, vol.~104,
  p.~195702, 2010.

\bibitem{araujo-2010}
N.~A.~M. Araújo and H.~J. Herrmann {\em Phys. Rev. Lett.}, vol.~105,
  p.~035701, 2010.

\bibitem{manna-2011}
S.~S. Manna and A.~Chatterjee {\em Physica A}, vol.~390, pp.~177--182, 2011.

\bibitem{shannon-1948}
C.~E. Shannon {\em Bell Syst. Tech. J.}, vol.~27, pp.~379--423, 1948.

\bibitem{fornberg-1988}
R.~Fornberg {\em Math. Comp.}, vol.~51, pp.~699--706, 1988.

\end{thebibliography}
\bibliographystyle{ieeetr}

\end{document}